\documentstyle[12pt]{article}
\begin{document}
\begin{center}
{\large \bf ENERGY DEPENDENCE OF RATIOS OF MULTIPLICITIES AND THEIR 
SLOPES FOR GLUON AND QUARK JETS}       \\
\bigskip
{\large  I.M. Dremin}\\
\bigskip
{\normalsize 
Lebedev Physical Institute, 117924 Moscow, Russia}
\end{center}
\bigskip

\begin{abstract}
The difference between the ratio of multiplicities and the ratio of 
their derivatives on energy for gluon and quark jets is calculated up to
next-to-next-to leading order of perturbative QCD. Its non-zero value is
uniquely defined by the running property of the QCD coupling constant.
It is shown that this difference is rather small compared to values which
can be obtained from experimental data. This 
disagreement can be ascribed either to strong non-perturbative terms or
to experimental problems with a scale choice, jets separation and inadequate 
assignement of soft particles to jets.
\end{abstract}
The ratio of multiplicities in gluon and quark jets has been of much debate
during last years. The lowest order prediction of the perturbative QCD drastically
overvalues experimental results for this ratio. Therefore, the higher-order
corrections up to the next-to-next-to leading terms of the perturbative QCD
have been calculated \cite{GM, DN}, and the ratio has been written as
\begin{equation}
\frac {\langle n_G\rangle }{\langle n_F\rangle }=r(y)=\frac {C_A}{C_F}(1-
r_{1}\gamma _{0}(y)-r_{2}\gamma _{0}^{2}(y))+O(\gamma _{0}^{3}),    \label{rat}
\end{equation}
where the scale of the process is defined by
$y=\ln Q/Q_{0}, \; Q$ is a virtuality of a jet, $\; Q_{0}$=const,
\begin{equation}
\gamma _{0}=\left (\frac {2C_{A}\alpha _{S}}{\pi }\right )^{1/2},
\end{equation}
$ \alpha _{S}$ is the running coupling 
constant, $\; C_{A}=3,\; C_{F}=4/3$ are Casimir operators, and
\begin{equation}
r_{1}=2\left [h_{1}+\frac {n_f}{12N_c}\left (1-\frac {2C_F}{N_c}\right )
\right ]-\frac {3}{4},                                               \label{r1}
\end{equation}
\begin{equation}
r_{2}=\frac {r_1}{6}\left (\frac {25}{8}-\frac {3}{4}\cdot \frac {n_f}{N_c}
-\frac {C_F}{N_c}\cdot \frac {n_f}{N_c}\right )+\frac {7}{8}-h_{2}-\frac {C_F}{N_c}
\cdot h_{3}+\frac {n_f}{12N_c}\cdot \frac {C_F}{N_c}h_4,   \label{r2}
\end{equation}
\begin{equation}
h_1=\frac {11}{24}, \, h_2=\frac {67-6\pi ^2}{36}, \, 
h_3=\frac {4\pi ^{2}-15}{24}, \, h_4= \frac {13}{3}. \label{h}
\end{equation}
Here, $n_f$ is the number of active flavours, $N_c$=3 the number of colours.
Let us note that only the first term of $r_2$ was obtained in \cite{GM} by
Feynman graphs technique. Other terms come from higher derivatives of the
generating functions \cite{DN} when the solution of the QCD equations for 
generating functions is found out by Taylor series expansion. They take into 
account the energy conservation in three-parton vertices.

Asymptotically, the ratio $r(y)$ (\ref{rat}) tends to a constant 
$C_{A}/C_F$=9/4 which shows that the gluon jet bremsstrahlung is stronger 
than for quark jets \cite{BG}. However, the correction terms are still noticeable
at presently accessible energies.

The energy dependence of average multiplicities of gluon and quark jets used
to be expressed in terms of the anomalous dimension $\gamma (y)$ as
\begin{equation}
\langle n_{G}(y)\rangle =Ke^{\int ^{y}\gamma (y')dy'}, \,\,
\langle n_{F}(y)\rangle =\langle n_{G}(y)\rangle /r(y).   \label{mu}
\end{equation}
Then it is easy to get the ratio of their derivatives as
\begin{equation}
\frac {\langle n_{G}(y)\rangle ' }{\langle n_{F}(y)\rangle '}=r(y)\left (
1+\frac {r'(y)}{\gamma (y)r(y)-r'(y)}\right ),   \label{der}
\end{equation}
or for its difference with the ratio of multiplicities one gets
\begin{equation}
D(y)=\frac {\langle n_{G}(y)\rangle '}{\langle n_{F}(y)\rangle '} -
\frac {\langle n_{G}(y)\rangle }{\langle n_{F}(y)\rangle }=
\frac {rr'}{\gamma r-r'}.   \label{diff}
\end{equation}
The anomalous dimension in terms of the running coupling constant looks like
\begin{equation}
\gamma =\gamma _{0}(1-a_{1}\gamma _{0})+O(\gamma _{0}^{3}).  \label{gam}
\end{equation}
Here
\begin{equation}
a_{1}=h_1 +\frac {n_f}{12N_c}\left (1-\frac {2C_F}{N_c}\right )-\frac {B}{2},\,
\, \; B=\frac {11N_{c}-2n_f}{24N_c}.   \label{a1B}
\end{equation}
The higher order terms have been omitted in our treatment. Taking into account
that
\begin{equation}
\gamma _{0}'= -h_{1}\gamma _{0}^{3} +O(\gamma _{0}^{5}),   \label{gapr}
\end{equation}
one estimates that the second term in the denominator of (\ref{diff}) 
can be neglected and gets finally
\begin{equation}
D(y)\approx \frac {r'(y)}{\gamma (y)} \approx \frac {N_c}{C_F}r_{1}h_{1}
\gamma _{0}^{2}(1+a_{1}\gamma _{0})
\left (1+\frac {2r_{2}\gamma _{0}}{r_1}\right ).    \label{Dy}
\end{equation}
The formulas (\ref{diff}) and (\ref{Dy}) are the main result of the paper.

It is important to stress that the difference $D(y)$ between the ratios of
derivatives and multiplicities directly demonstrates the running property
of the QCD coupling constant. It is identically equal to zero for fixed 
coupling because $r'=0$ in that case. For the running coupling, this difference
is always positive and proportional to the value of the coupling constant.
Consequently, it tends to zero in asymptotics.

The corrections due to the anomalous dimension (\ref{gam}) and due to the
multiplicity ratio (\ref{rat}) have been specially left as separate factors in 
brackets in (\ref{Dy}) to show their relative importance. One easily notices 
that the last correction in $D(y)$ is stronger than in the ratio (\ref{rat})
itself because $r'$ in the numerators of (\ref{diff}) and (\ref{Dy}) acquires
the factors $n$ when differentiating the subsequent terms of $\gamma _{0}^{n}$
in (\ref{rat}).

Asymptotically the terms in brackets tend to 1 since $\gamma _{0} \rightarrow
0$, but at present energies ($\gamma _{0}\approx 0.45 -0.5$ at $Z^0$) the
corrections are rather large. In the energy range near $Z^0$, the factor in 
front of the brackets in (\ref{Dy}) is about 0.05--0.06. The anomalous 
dimension correction in the first bracket enlarges the value of $D$ by about
15$\%$. More important is the expression in the second bracket. With values of
$r_1$ and $r_2$ given by (\ref{r1}), (\ref{r2}) and inserted in (\ref{Dy}),
one estimates it as about 3.2, i.e. the linear in $\gamma _0$ term (NNLO
correction) contributes more than 1 (MLLA term), and therefore next terms
should be evaluated. Anyway taking this expression for granted, one would
estimate $D$ as
\begin{equation}
D\approx 0.16 - 0.20.   \label{est}
\end{equation}
Thus the predicted value of $D$ is comparatively small.

Let us note that the first term from (\ref{r2}) only as calculated first
in \cite{GM} contributes to the second bracket the value about 0.25, and
the corresponding contribution to $D$ is equal approximately 0.06 - 0.08.

We have considered the one-scale problem without imposing any limitations
on the transverse momentum and considering the jet evolution with energy
in the total phase space. In experiment, the more severe restrictions are 
often introduced due to specifics of installations or due to some special
criteria. In particular, one can consider jet evolution as a function of 
some internal scale (e.g., similar to the jet transverse momentum) at a 
given energy (usually chosen at $Z^0$ peak because of larger statistics).
Such a procedure has been exploited in Refs \cite{AL, DE}. It was claimed
that the proportionality of the two scales favours comparison of theoretical 
predictions with experimental data on the evolution of above ratios. The scale
chosen in \cite{DE} was
\begin{equation}
\kappa =E_{jet}\sin \frac {\theta }{2},     \label{ka}    
\end{equation}
where $E_{jet}$ is the energy of the jet and $\theta $ the angle with respect 
to the closest jet. In \cite{AL}, the similar scale of the geometric mean of
such scales of the gluon jet with respect to both quark jets in three-jet
events has been used. From the data presented in \cite{DE} one can estimate 
that the difference $D(\kappa )$ ranges approximately from 0.95 at $\kappa $=
7 GeV to 0.7 at $\kappa $=28 GeV. These values are much larger than predicted 
above.

However, both theoretical and experimental approaches to the problem should 
be further analysed. The large correction in Eq. (\ref{Dy}) implies that 
next terms of the perturbative expansion can be essential in the ratio of
derivatives. They are enlarged in (\ref{Dy}) as discussed above and support
earlier conclusions (see, e.g., Ref. \cite{DR}) that subleading effects are
quantitatively important at experimentally accessible energies. Our experience
tells us also that power corrections due to conservation laws related to recoil
effects can become essential as well \cite{LU}.

Experimental procedures raise even more questions. There exists the scale 
dependence in the final result if one compares the data of \cite{AL} with that
of \cite{DE}. The problem of the scale choice has recently been discussed in
Ref. \cite{ED}. Besides, it has been shown \cite{ED} that different algorithms
of the jet selection and ascribing soft particles to a jet give rise to
somewhat different conclusions concerning jet multiplicities and their slopes.
Further study is necessary.

Another problem raised by the procedure used in Ref. \cite{DE} is the role of 
non-perturbative effects in average multiplicities of gluon and quark jets.
It has been claimed \cite{DE} that the hadronization of quark jets leads to
some constant (an assumed ansatz!) excess over gluon jets
equal to about two or three particles. It compensates somewhat the 
perturbative  QCD excess of the gluon jet multiplicities. This constant term
in energy dependence of multiplicities has been ascribed to a possible
non-perturbative contribution. Unfortunately, this statement can hardly be
confronted to any reliable theoretical treatment. These terms do not contribute
to derivatives of the multiplicities but drastically diminish the multiplicity
ratio thus increasing their difference $D(y)$.

At the same time, the theoretical estimate (\ref{est}) seems quite reasonable
qualitatively if one accepts the values of the ratio of derivatives as about
2.1 given in \cite{DE} and the value of the ratio of multiplicities about
1.84 predicted theoretically \cite{DR} with the experimental result $\sim 1.6$
claimed in \cite{OP}.

In conclusion, we have evaluated the difference between the ratio of slopes of 
average multiplicities and the ratio of multiplicities in gluon and quark jets
in the next-to-next-to leading approximation of the perturbative QCD and 
confronted it to some experimental data.
\vspace{1mm}

ACKNOWLEDGEMENTS

This work was supported in part by the Russian Fund
for Basic Research (grant 96-02-16347) and by INTAS.

\end{document}